\documentclass[12pt]{article}
\usepackage{graphicx}
\graphicspath{{ps/}}

\newcommand\pubnumber{}
\newcommand\pubdate{\today}

\def\Title#1{\begin{center} {\Large #1 } \end{center}}
\def\Author#1{\begin{center}{ \sc #1} \end{center}}
\def\Address#1{\begin{center}{ \it #1} \end{center}}

\newcommand\pubblock{\rightline{\begin{tabular}{l} \pubnumber\\
         \pubdate  \end{tabular}}}
\newenvironment{Abstract}{\begin{quotation}  }{\end{quotation}}
\newenvironment{Presented}{\begin{quotation} \begin{center} 
             PRESENTED AT\end{center}\bigskip 
      \begin{center}\begin{large}}{\end{large}\end{center} \end{quotation}}

\begin{document}
\begin{titlepage}
\pubblock

\vfill
\Title{Measurements of $\sin2\phi_1$ in $B^0\to\eta^\prime K^0$,
  $\omega K_S^0$ and $\pi^0 K^0$ decays}
\vfill
\Author{Tagir Aushev}
\Address{Institute for Theoretical and Experimental Physics, Moscow, Russia \\ 
aushev@itep.ru}
\vfill
\begin{Abstract}
In this report we summarize the most recent $\sin2\phi_1$ measurements 
in the $b\to q\bar qs$ decays.
\end{Abstract}
\vfill
\begin{Presented}
6th International Workshop on the CKM Unitarity Triangle\\
University of Warwick, Great Britain, September 6-10, 2010
\end{Presented}
\vfill
\end{titlepage}
\def\thefootnote{\fnsymbol{footnote}}
\setcounter{footnote}{0}

\section{Introduction}

Decays of $B$ mesons mediated by $b\to s$ penguin amplitudes play an
important role in both measuring the Standard Model (SM) parameters 
and in probing new physics. In the decay 
$\Upsilon(4S)\to B^0\bar B^0\to f_{CP}f_{\rm tag}$, where one of the 
$B$ mesons decays at time $t_{CP}$ to a $CP$ eigenstate $f_{CP}$ and
the other decays at time $t_{\rm tag}$ to a final state $f_{\rm tag}$
that distinguishes between $B^0$ and $\bar B^0$, the decay rate has a
time dependence given by
\begin{equation}
{\cal P}(\Delta t)=\frac{e^{-|\Delta t|/\tau_{B^0}}}{4\tau_{B^0}}
\left[ 1+q\cdot[{\cal S}_f\sin(\Delta m_d\Delta t)+
{\cal A}_f\cos(\Delta m_d\Delta t)]\right].
\end{equation}
Here, ${\cal S}_f$ and ${\cal A}_f$ are parameters that describe 
mixing-induced and direct $CP$ violation, respectively, $\tau_{B^0}$
is the $B^0$ lifetime, $\Delta m_d$ is the mass difference between the
two $B^0$ mass eigenstates, $\Delta t=t_{CP}-t_{\rm tag}$, and the
$b$-flavor charge, $q=+1(-1)$ when the tagged $B$ meson is a 
$B^0(\bar B^0)$.
The SM predicts ${\cal S}_f=\xi_f\sin2\phi_1$ ($\phi_1\equiv\beta$) 
and ${\cal A}_f\approx0$ to a good approximation for most of the 
decays that proceed via $b\to sq\bar q$ ($q=c,s,d,u$) quark 
transitions~\cite{grossman}, where $\xi_f=+1(-1)$ corresponds to 
$CP$-even(-odd) final states and $\phi_1$ is an angle of the 
unitary triangle.
However, even within the SM both ${\cal S}_f$ and ${\cal A}_f$ could
be shifted due to the contribution of a color-suppressed tree diagram 
that has a $V_{ub}$ coupling~\cite{zupan}.  Recent SM 
calculations~\cite{beneke} for the effective $\sin2\phi_1$ values,
$\sin2\phi_1^{\rm eff}$, for $B^0\to\eta^\prime K^0$, $\omega K_S^0$ 
and $\pi^0 K^0$ decay modes agree with $\sin2\phi_1$, as measured 
in $B^0\to J/\psi K^0$ decays, at the level of $0.1-0.01$ depending 
on the decay mode.  Thus a comparison of the ${\cal S}_f$ and 
${\cal A}_f$ measurements  between modes and a search for larger
deviations of ${\cal S}_f$ from $\sin2\phi_1$ is an important test of 
the SM.

Here we present the recent measurements of mixing-induced $CP$
violation in $B^0$ decays into $\eta^\prime K^0$, $\omega K_S^0$ 
and $\pi^0 K^0$ final states.  These results are obtained by the two 
experiments, BaBar and Belle, working at the 
energy-asymmetric $e^+e^-$ colliders at the $\Upsilon(4S)$ resonance.
The statistics used by BaBar is $467\times10^6\,B\bar B$ pairs.  The 
Belle analyses of $B^0\to\eta^\prime K^0$ and $B^0\to\omega K_S^0$ use 
$535\times10^6\,B\bar B$ pairs and $B^0\to\pi^0 K^0$ is based on the 
statistics of $657\times10^6\,B\bar B$ pairs.

The $B$ candidates are identified using the energy difference 
$\Delta E=E_B-E_{\rm beam}$ and beam-energy constrained mass 
$M_{\rm bc}=\sqrt{E_{\rm beam}^2-p_B^{*2}}$, where $E_B$ and $p_B^*$ 
are the $B$ candidate energy and momentum in the center-of-mass (CM) 
system, respectively.  In case of $K_L$ in the decay chain, only the
direction of the $K_L$ momentum is used and a kinematical constraint
to the $B$ mass.  In this case $\Delta E$ is
used in BaBar analyses and $p_B^*$ in Belle to identify the signal.

The dominant background for the $b\to s\bar qq$ signal comes from
continuum events $e^+e^-\to q\bar q$ where $q=u,d,s,c$.  To 
distinguish these topologically jet-like events from the spherical $B$ 
decay signal events, a set of variables that characterize the event 
topology are combined into a signal (background) likelihood variable 
${\cal L}_{\rm sig}$ (${\cal L}_{\rm bkg}$) (Belle) or into a Neural 
Network (BaBar).

Since the $B^0$ and $\bar B^0$ mesons are approximately at rest in 
the $\Upsilon(4S)$ center-of-mass (CM) system, $\Delta t$ can be 
determined from the displacement in $z$ between the $f_{CP}$ and 
$f_{\rm tag}$ decay vertices: $\Delta t\approx(z_{CP}-z_{\rm tag})/
(\beta\gamma c)\equiv\Delta z/(\beta\gamma c)$.

\section{Results}

\begin{figure}[tb]
\centering
\includegraphics[height=2.6in]{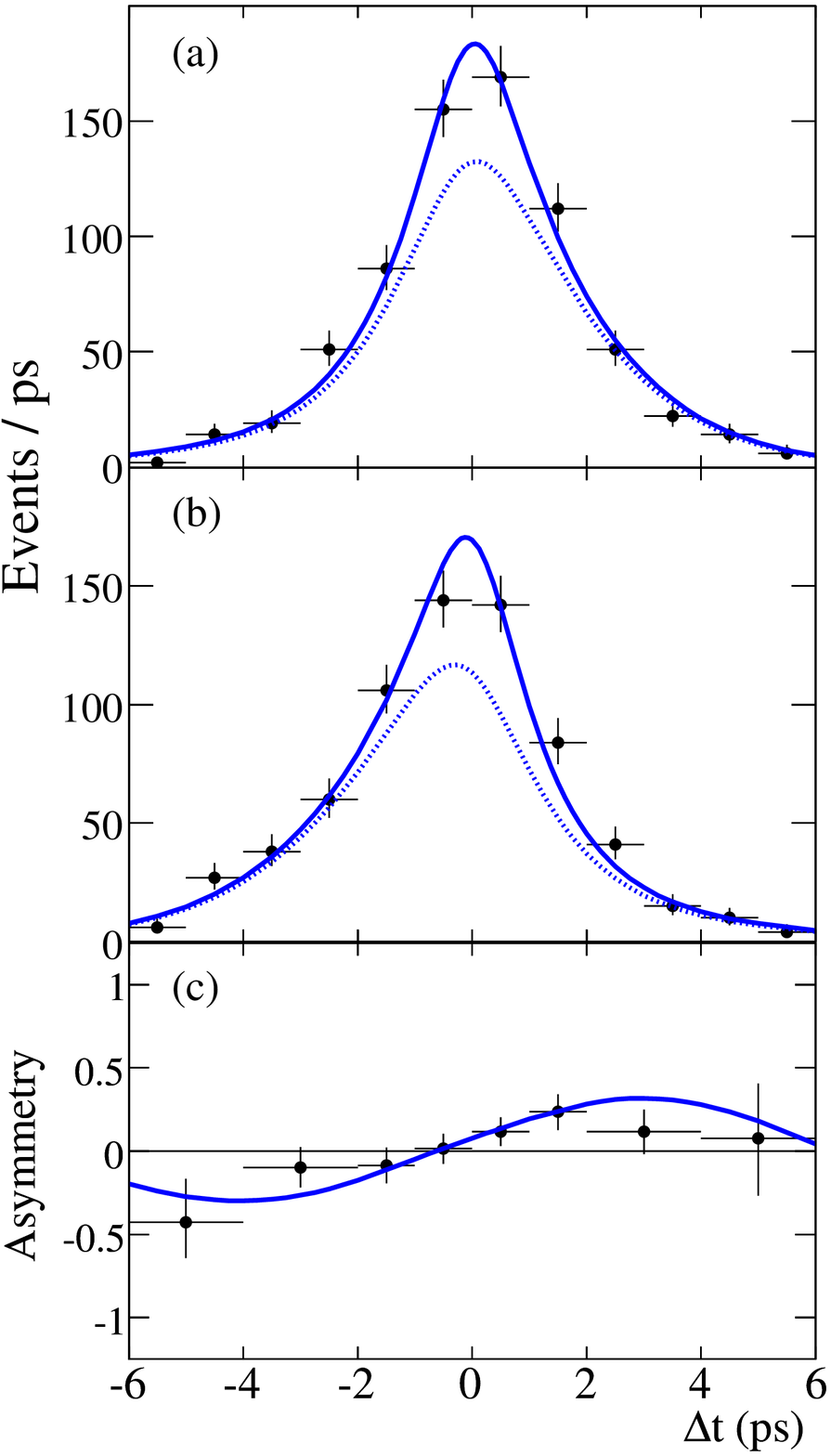}
\includegraphics[height=2.6in]{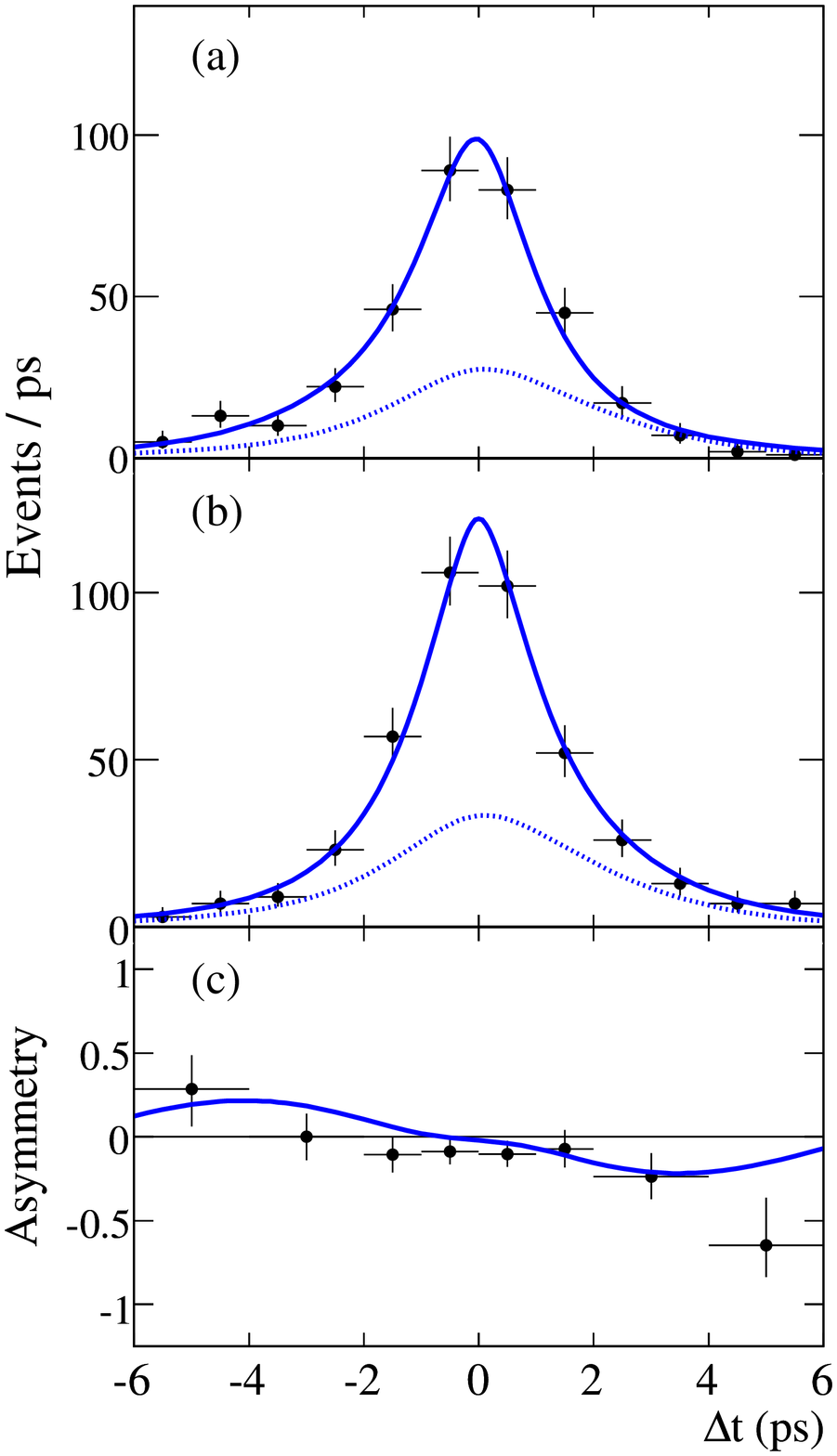}
\includegraphics[height=2.6in]{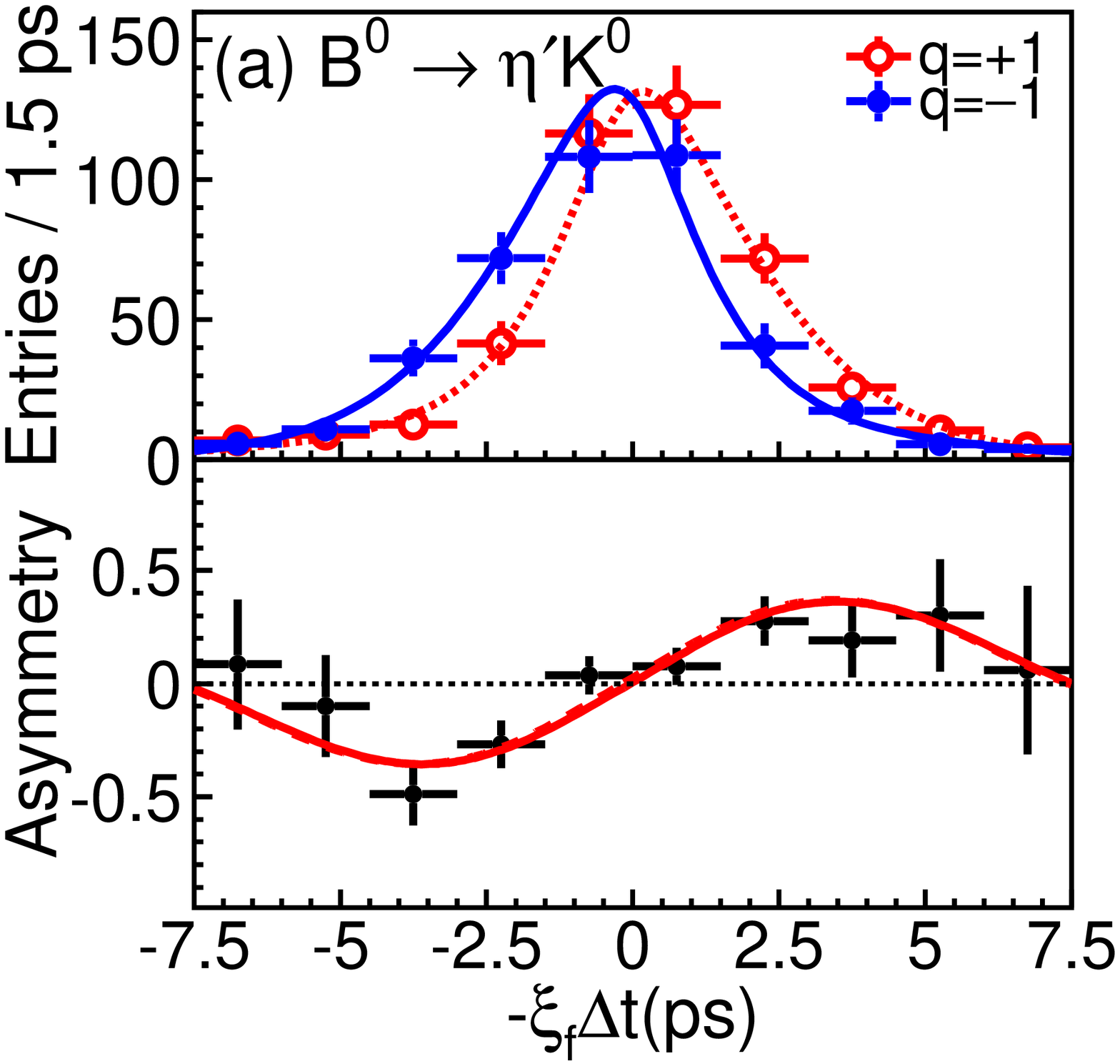}
\caption{Distribution of $\Delta t$ and of raw asymmetry as a function
of $\Delta t$, with fit results overlaid, for $B^0\to\eta^\prime K^0$
events.  The left and central plots represent BaBar results for 
$K_S^0$ and $K_L^0$, respectively, subdivided in
(a) $B^0$ tags and (b) $\bar B^0$ tags; the solid (dotted) line 
displays the total (signal) fit function;
the right plots show the Belle distributions for well tagged, combined
$K_S^0$ and $K_L^0$ events.}
\label{fig:etaK0}
\end{figure}

In both experiments the same decay modes are used to reconstruct 
$B_{CP}\to\eta^\prime K^0$ candidates: 
$B^0\to\eta^\prime(\rho\gamma$, $\eta_{\gamma\gamma}\pi^+\pi^-$,
$\eta_{3\pi}\pi^+\pi^-)$ $K_S^0(\pi^+\pi^-)$, 
$\eta^\prime(\rho\gamma$, $\eta_{\gamma\gamma}\pi^+\pi^-)$
$K_S^0(\pi^0\pi^0)$ and $\eta^\prime(\eta_{\gamma\gamma}\pi^+\pi^-$,
$\eta_{3\pi}\pi^+\pi^-)$ $K_L^0$.  Significant signals are observed in
all channels~\cite{babar_all,belle_eta}.  For the reconstructed events
the ${\cal S}_f$ and ${\cal C}_f$ are determined by performing an 
unbinned maximum-likelihood fit to the $\Delta t$ distributions.  The 
$\Delta t$ distributions with the superimposed results of the fits are 
presented on Fig.~\ref{fig:etaK0}.  The obtained signal yields and 
results of the $\Delta t$ fits are summarized in 
Table~\ref{tab:results}.  The decay $B^0\to\eta^\prime K^0$ is the 
only $b\to s$ mode where a significant $CP$ violation is measured.

\begin{figure}[htb]
\centering
\includegraphics[height=1.5in]{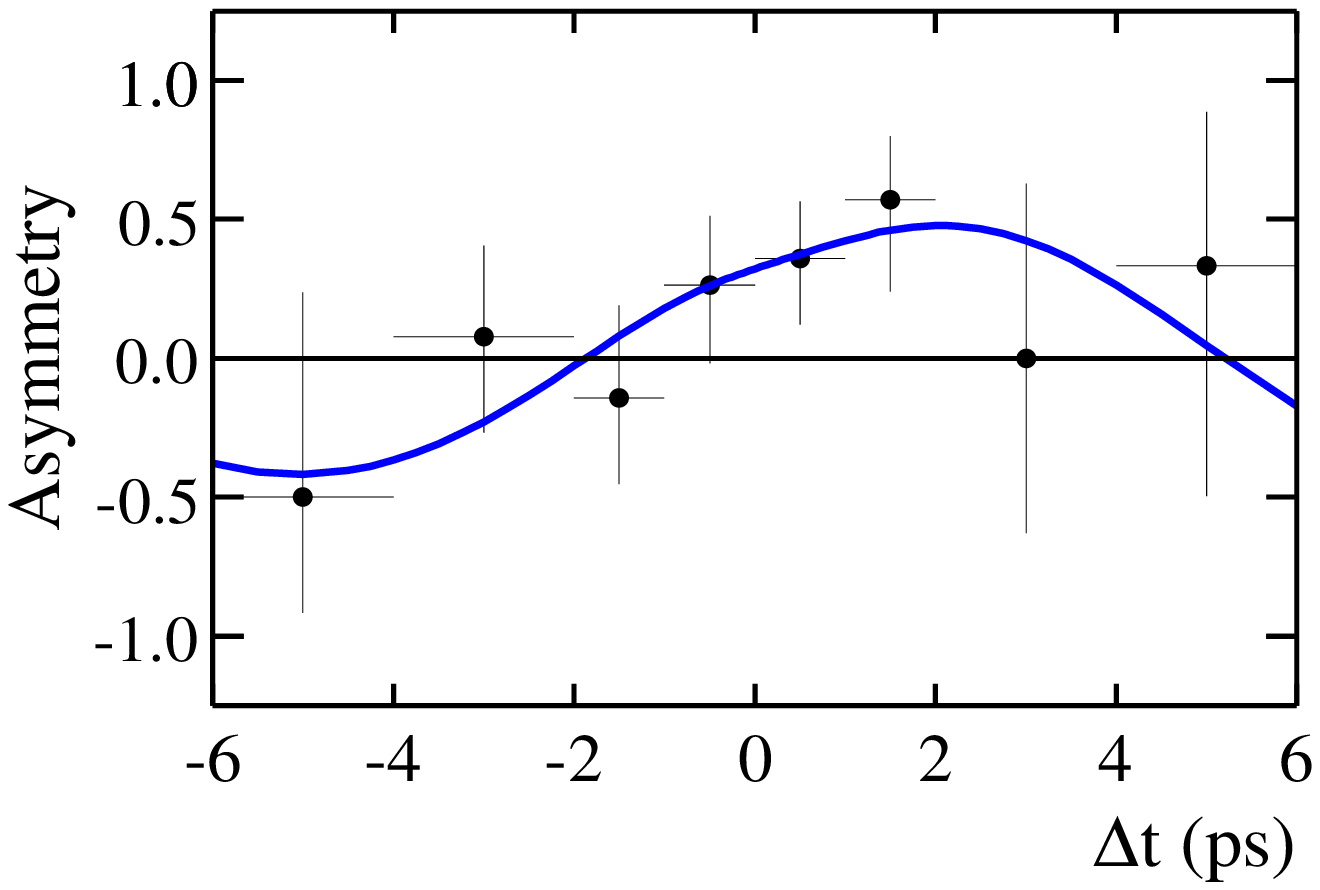}
\includegraphics[height=1.5in]{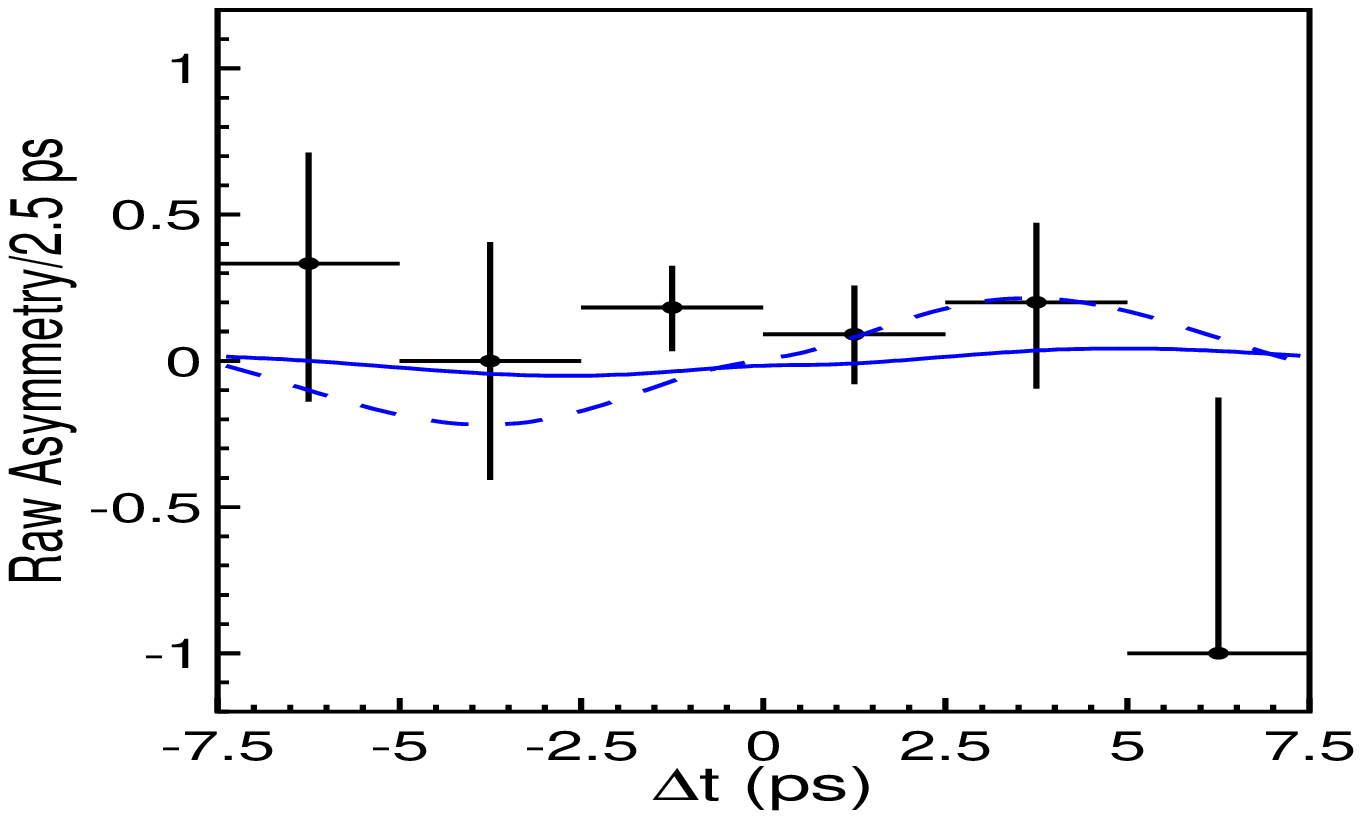}\\
\includegraphics[height=1.5in]{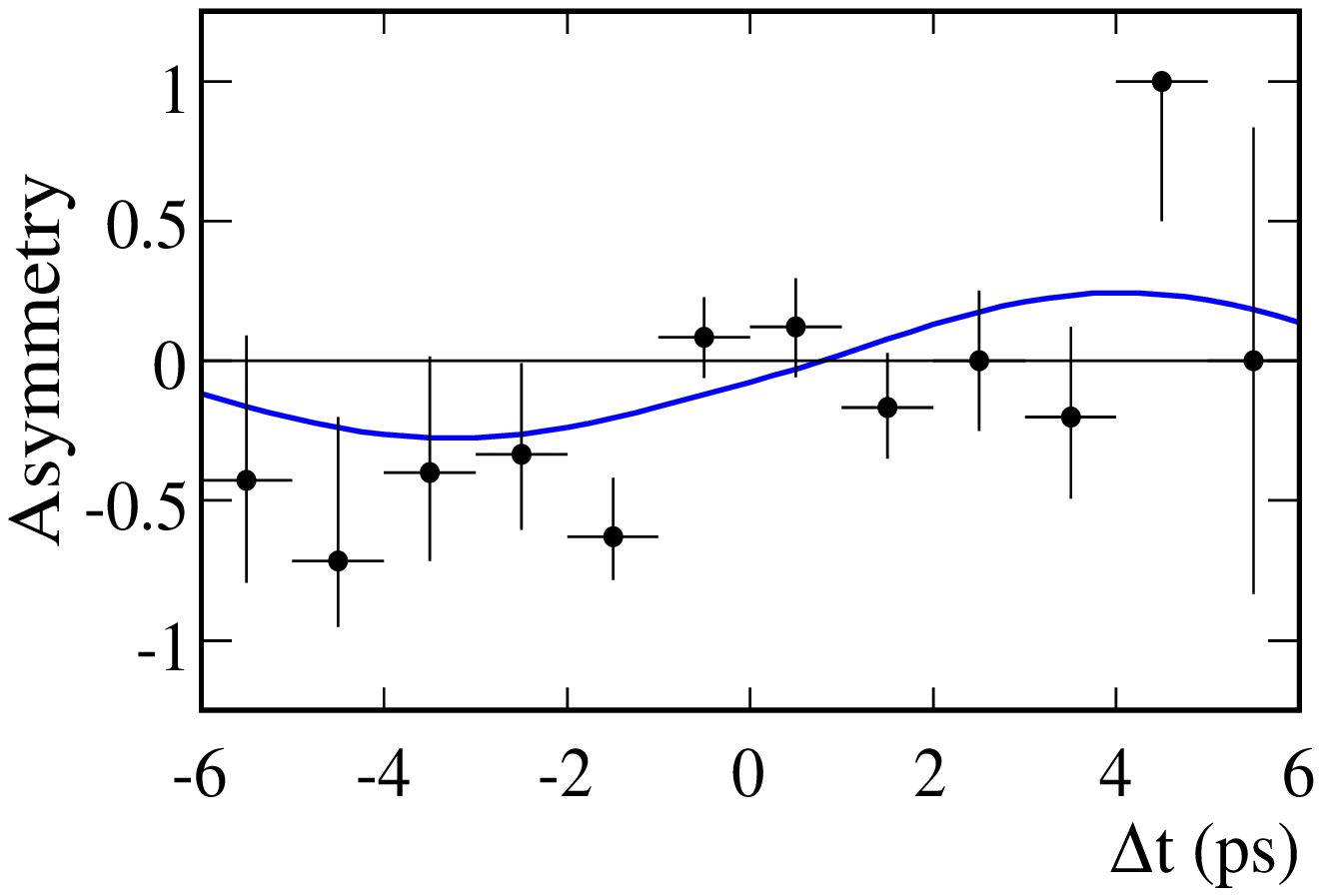}
\includegraphics[height=1.5in]{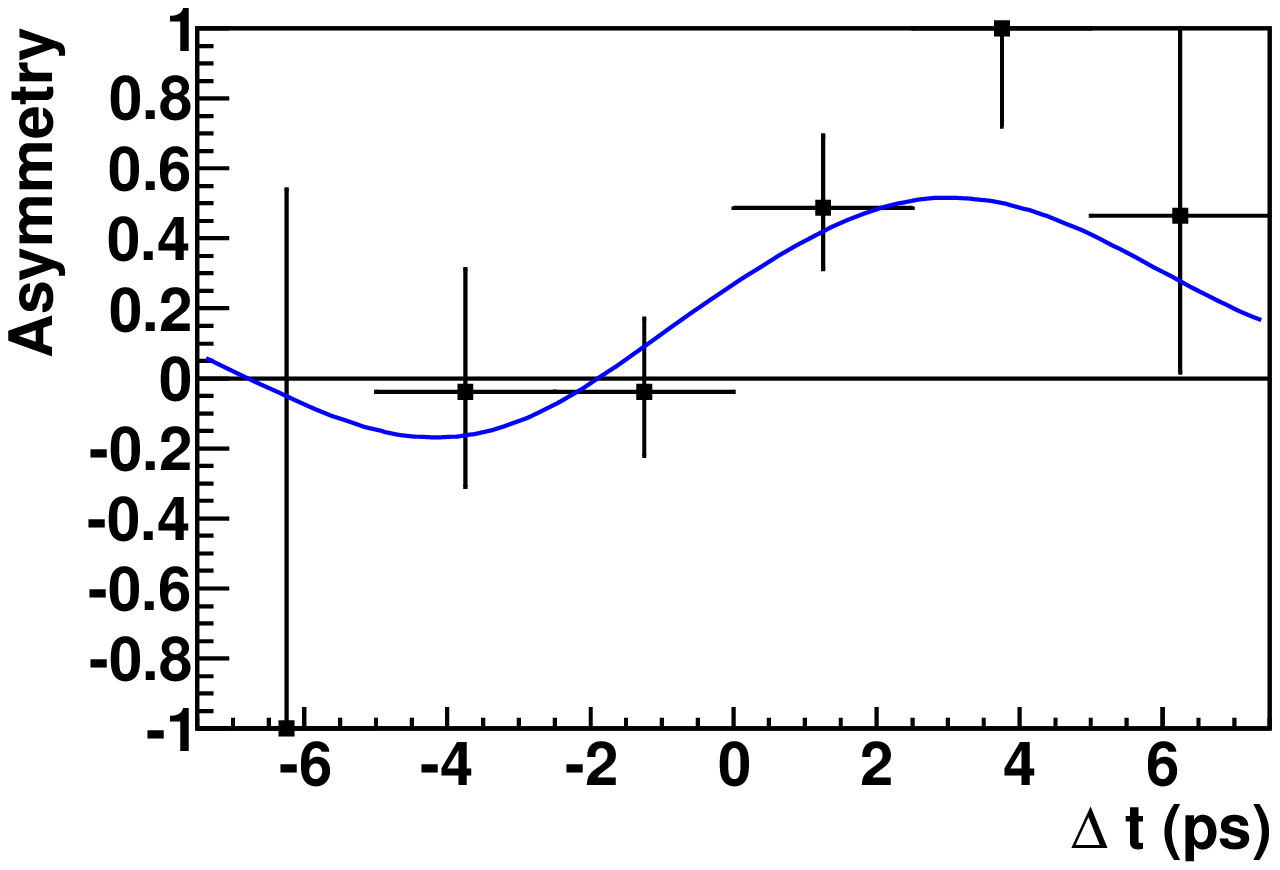}
\caption{Asymmetries of good-tagged events:
(top) $B^0\to\omega K_S^0$; (bottom) $B^0\to K^0\pi^0$; 
(left) BaBar; (right) Belle.  The solid curves show the results of 
the fits, the dashed line shows the SM expectation.}
\label{fig:omega_pi0_K0}
\end{figure}

Similar analyses are performed for the $B^0\to\omega K_S^0$ decay 
mode~\cite{babar_all,belle_omega}.  The results are included in 
Table~\ref{tab:results} and the $\Delta t$ symmetries are presented in
Fig.~\ref{fig:omega_pi0_K0}.

For the $B^0\to\pi^0 K^0$ decay mode ${\cal A}_{K^0\pi^0}$ can be
predicted within a few percent precision by applying an isospin sum 
rule to the recent measurements of $B$ decays into $K\pi$ final 
states~\cite{sumrule}:
\begin{equation}
{\cal A}_{K^+\pi^-}+{\cal A}_{K^0\pi^+}
\frac{{\cal B}(K^0\pi^+)\tau_{B^0}}{{\cal B}(K^+\pi^-)\tau_{B^+}}=
{\cal A}_{K^+\pi^0}
\frac{2{\cal B}(K^+\pi^0)\tau_{B^0}}{{\cal B}(K^+\pi^-)\tau_{B^+}}+
{\cal A}_{K^0\pi^0}
\frac{2{\cal B}(K^0\pi^0)}{{\cal B}(K^+\pi^-)}.
\end{equation}
A significant discrepancy between the measured and expected values 
of ${\cal A}_{K^0\pi^0}$ would indicate a new physics contribution to
the sum rule.  Both experiments provided measurements of $CP$ 
violating parameters in this decay mode, Belle using $K_S^0\pi^0$ and 
$K_L^0\pi^0$ modes and BaBar using $K_S^0\pi^0$ 
only~\cite{babar_all,belle_pi0}.  The results are summarized in 
Table~\ref{tab:results}, the $\Delta t$ asymmetries are presented in 
Fig.~\ref{fig:omega_pi0_K0}.

\begin{table}[t]
\begin{center}
\begin{tabular}{clccc}
Mode&& Signal yield & $-\eta_f{\cal S}_f$ & ${\cal C}_f=-{\cal A}_f$\\
\hline
$\eta^\prime K_S^0$
&BaBar & $1457\pm43$ & $+0.53\pm0.08\pm0.02$ & $-0.11\pm0.06\pm0.02$\\
&Belle & $1421\pm46$ & $+0.67\pm0.11$ & $+0.03\pm0.07$ \\
\hline
$\eta^\prime K_L^0$
&BaBar &  $416\pm29$ & 
$+0.82^{+0.17}_{-0.19}\pm0.02$ & $+0.09^{+0.13}_{-0.14}\pm0.02$\\
&Belle &  $454\pm39$ & $+0.46\pm0.24$ & $-0.09\pm0.16$ \\
\hline
$\eta^\prime K^0$
&BaBar & & $+0.57\pm0.08\pm0.02$ & $-0.08\pm0.06\pm0.02$ \\
&Belle & & $+0.64\pm0.10\pm0.04$ & $+0.01\pm0.07\pm0.05$ \\
\hline
$\omega K_S^0$
&BaBar & $121\pm13$ & 
$+0.55^{+0.26}_{-0.29}\pm0.02$ & $-0.52^{+0.22}_{-0.20}\pm0.03$ \\
&Belle & $118\pm18$ & $+0.11\pm0.46\pm0.07$ & $+0.09\pm0.29\pm0.06$ \\
\hline
$K_S^0\pi^0$
&BaBar & $411\pm24$ & 
$+0.55\pm0.20\pm0.03$ & $+0.13\pm0.13\pm0.03$ \\
$K_S^0\pi^0$
&Belle & $634\pm34$ & & \\
$K_L^0\pi^0$
&Belle & $285\pm52\,(3.7\sigma)$ & & \\
$K^0\pi^0$
&Belle & & 
$+0.67\pm0.31\pm0.08$ & $-0.14\pm0.13\pm0.06$ \\
\end{tabular}
\caption{The signal yields and the results of the fits to the 
$\Delta t$ distributions, the first errors are statistical and the 
second errors are systematic.}
\label{tab:results}
\end{center}
\end{table}

\section{Summary}

The current results of the $CP$ violation measurements in the decay 
modes $B^0\to\eta^\prime K^0$, $\omega K_S^0$ and $\pi^0 K^0$ are 
consistent with SM.  Although further updates are expected from the 
Belle experiment using the whole data statistics, more significant 
improvements can be provided in the future by Super $B$-factories.

\end{document}